\documentclass[conference]{IEEEtran}
\IEEEoverridecommandlockouts

\usepackage{cite}
\usepackage{multirow}
\usepackage{amsmath,amssymb,amsfonts}
\usepackage{graphicx}
\usepackage{textcomp}
\usepackage{xcolor}
\usepackage{booktabs}
\usepackage{siunitx}
\usepackage[hidelinks]{hyperref}
\usepackage{algorithm}
\usepackage{algorithmic}
\usepackage{eso-pic}

\usepackage{color}
\usepackage{soul}

\newcommand{\dBm}{\,\text{dBm}}
\newcommand{\GHz}{\,\text{GHz}}

\newcommand{\GP}{\mathcal{GP}}

\newcommand{\stageset}{\mathcal{S}}

\def\BibTeX{{\rm B\kern-.05em{\sc i\kern-.025em b}\kern-.08em
    T\kern-.1667em\lower.7ex\hbox{E}\kern-.125emX}}

\begin{document}

\AddToShipoutPictureFG*{%
  \AtPageLowerLeft{%
    \hspace{0.75in}\raisebox{0.55in}{%
      \begin{minipage}{6.5in}
      \scriptsize\noindent
      \copyright\ 2026 IEEE. Personal use of this material is permitted.
      Permission from IEEE must be obtained for all other uses, in any
      current or future media, including reprinting/republishing this
      material for advertising or promotional purposes, creating new
      collective works, for resale or redistribution to servers or
      lists, or reuse of any copyrighted component of this work in
      other works.
      \end{minipage}%
    }%
  }%
}

\title{Robust Base Station Placement in Agricultural IoT via Bayesian Optimization}

\author{
  \IEEEauthorblockN{Gourav Prateek Sharma\IEEEauthorrefmark{1}, 
                    Durgesh Singh\IEEEauthorrefmark{2}, 
                    James Gross\IEEEauthorrefmark{3}}
  \IEEEauthorblockA{\IEEEauthorrefmark{1}Dept.\ of ECE, National Institute of Technology Kurukshetra, India, 
                    gourav.sharma@nitkkr.ac.in}
  \IEEEauthorblockA{\IEEEauthorrefmark{2}Dept.\ of ECE, Thapar Institute of 
                    Engineering \& Technology, India, durgesh.singh@thapar.edu}
  \IEEEauthorblockA{\IEEEauthorrefmark{3}School of EECS, KTH Royal Institute of 
                    Technology, Sweden, jamesgr@kth.se}
}

\maketitle

\begin{abstract}
Precision-agriculture networks based on private 5G NR should ensure reliable connectivity for IoT sensor nodes throughout the crop growing season, yet the propagation environment changes dramatically as vegetation grows and matures. We formulate $K$-base-station~(BS) placement as a \textit{maximin seasonal coverage} problem that maximizes the worst-case coverage fraction across all crop growth stages. Since each objective evaluation requires expensive ray-tracing simulations across all stages, we adopt a Gaussian-process Bayesian optimization~(GPBO) framework that builds a probabilistic surrogate of the robust objective using ray tracing. On a $1\,\text{km}^2$ multi-crop farm with three distinct crop zones at $3.5\,\text{GHz}$, the proposed scheme achieves $72.8\%$ worst-case coverage with $K{=}3$ BSs in fewer than fifty ray-tracing evaluations, outperforming budget-matched state-of-the-art approaches by at least $4.6\,\text{pp}$ across all four seasonal stages. 
\end{abstract}


\section{Introduction}
\label{sec:intro}

Precision agriculture relies on a wide range of IoT sensors to monitor parameters such as soil moisture, crop health, and irrigation flow. These sensors are typically deployed in dense and irregular patterns across farmland, alongside mobile agricultural machinery that also requires connectivity. While sub-GHz LPWAN technologies such as LoRa and NB-IoT currently dominate low-data-rate agricultural sensing, emerging use cases including autonomous machinery telemetry and drone-assisted crop scouting demand higher throughput and lower latency, motivating the adoption of private 5G networks for precision agriculture \cite{majumdar2025enhancing}. Ensuring reliable communication for all devices within the farm boundary is therefore essential \cite{elijah2018overview}. However, deploying communication infrastructure in rural areas is often limited by cost, necessitating the use of only a few base stations (BSs) to cover large agricultural regions \cite{majone2013wireless}.

A major challenge in such environments is the dynamic nature of the radio propagation channel, which changes significantly over the crop growth cycle \cite{garcia2021deployment}. As vegetation develops, factors such as canopy height, water content, and leaf density increase, altering the electromagnetic characteristics of the environment \cite{rahim2017foliage}. These variations lead to changes in signal attenuation and scattering, causing coverage patterns to evolve throughout the season. Consequently, a BS placement strategy that is effective early in the season may suffer considerable performance degradation as crops mature. Designing BS deployments that remain robust under these seasonal variations is therefore a critical challenge for agricultural IoT networks.

Base station placement has been widely studied in wireless network planning \cite{hurley2002planning}. Classical approaches rely on simplified propagation models and search strategies such as metaheuristics based on Particle Swarm Optimization (PSO) and genetic algorithms \cite{yangyang2004particle,valavanis2014base}. More recently, Bayesian optimization (BO) has emerged as an effective technique for optimizing expensive black-box objectives such as network coverage or capacity. BO has been applied to problems including wireless network planning, antenna configuration and transmitter placement
in complex propagation environments \cite{wu2024multiobjective,sato2024bayesian}. In parallel, modern ray-tracing tools (e.g., Sionna RT \cite{hoydis2023sionna}) enable physically accurate modelling of radio propagation in detailed environments, making it possible to evaluate coverage under realistic geometry and material properties. Existing BS placement optimization studies typically focus on a static propagation environment, especially in urban scenarios. However, in agricultural deployments, vegetation growth introduces substantial variation in channel conditions over the span of weeks and months. Optimizing BS locations for a single propagation snapshot can therefore lead to suboptimal performance over the full growing season. For instance, a placement that maximizes line-of-sight links over bare soil in spring may suffer severe signal blockage and network partitioning when crop canopies reach maximum height and foliage density in late summer.
To the best of our knowledge, the problem of \emph{seasonal-robust BS placement} in agricultural IoT environments using physically based propagation models has not been studied. In this work, we address this problem by formulating BS placement as a maximin seasonal coverage optimization task. The objective is to maximize the worst-case coverage fraction across multiple crop growth stages. As evaluating coverage requires large-scale measurements or computationally expensive ray-tracing simulations, we employ Gaussian-process Bayesian optimization (GPBO) to efficiently search the placement space.

The main contributions of this paper are as follows:
\begin{itemize}
    \item We introduce a seasonal channel model for agricultural 
    wireless networks based on ITU-R P.833~\cite{itu_p833} and formulate 
    BS placement as a maximin problem that maximizes the worst-case 
    coverage fraction across all crop growth stages.
    \item We propose a GPBO framework that efficiently solves the 
    placement problem using a limited number of ray-tracing evaluations 
    and demonstrate its superiority over budget-matched state-of-the-art 
    approaches across all seasonal stages.
\end{itemize}





The rest of the paper is organized as follows. Section~\ref{sec:sysmod} describes the farm scene, seasonal channel model and formulation of the seasonal-robust BS placement problem. Section~\ref{sec:bo} presents proposed Bayesian optimization framework. Section~\ref{sec:results} discusses the ray tracing setup and the evaluation of the proposed method with respect to the state-of-the-art, and Section~\ref{sec:conclusion} concludes the article and discusses future research directions.   
\section{System Model and Problem Formulation}
\label{sec:sysmod}

\subsection{Farm Scene and Geometry}

\begin{figure}[t]
  \centering
  \includegraphics[trim={1cm 3.85cm 1cm 3.85cm},width=\columnwidth]{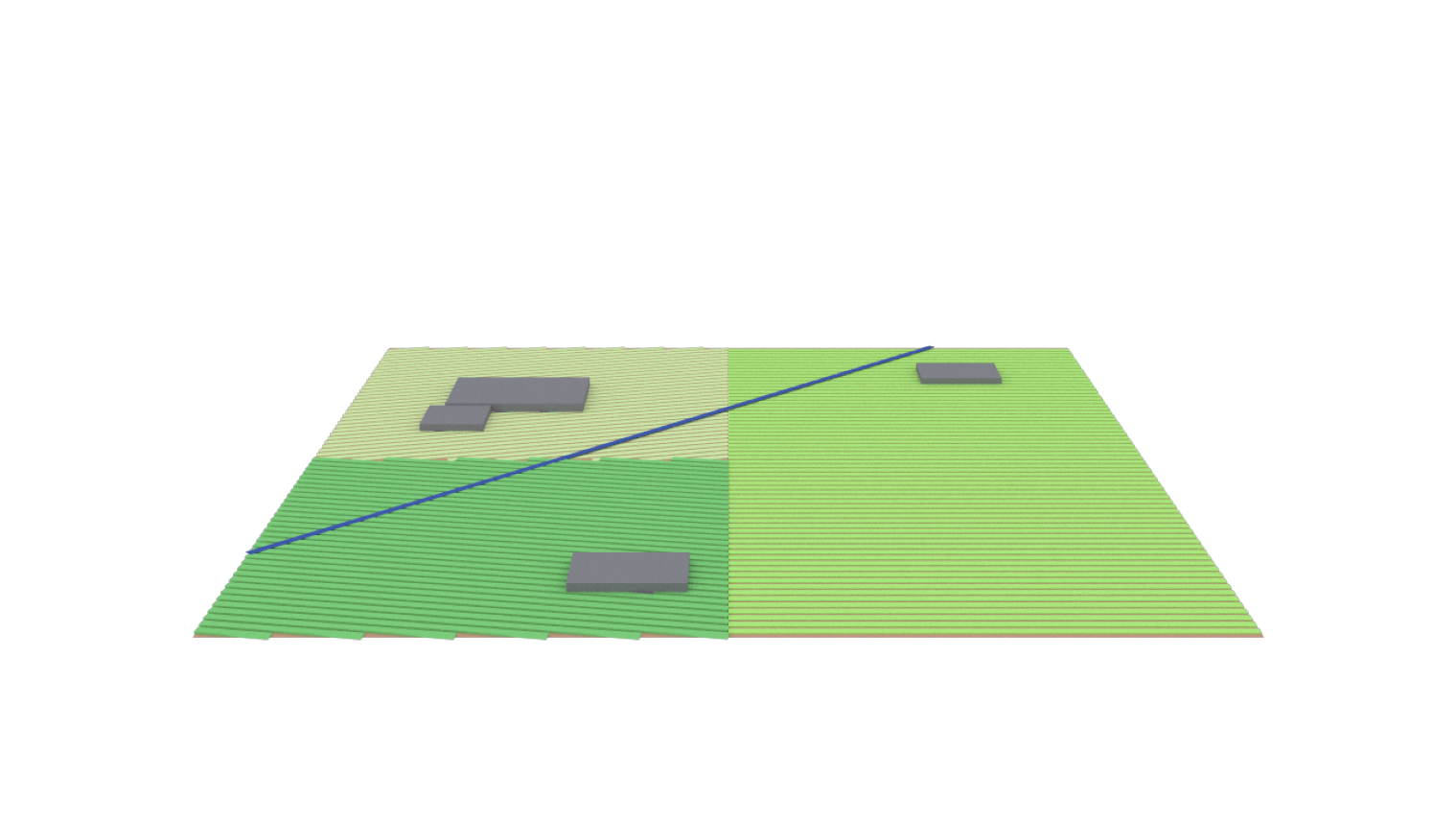}
  \caption{ 3D illustration of the farm scene considered in the problem.}
  \label{fig:scene_preview}
\end{figure}

The farm is a flat field \SI{1}{\km\squared} centred at the origin, $x, y \in [-500, 500]\,\text{m}$, divided into three crop zones with different seasonal dynamics  (Fig.~\ref{fig:scene_preview}):
\begin{itemize}
  \item \emph{Crop1} (NW quadrant, $x \in [-500,0]$, $y \in [0,500]\,\text{m}$):
        rows at $15^\circ$ with respect to E--W.
  \item \emph{Crop2} (SW quadrant, $x \in [-500,0]$, $y \in [-500,0]\,\text{m}$):
        rows at $-10^\circ$ with respect to E--W; tallest and most attenuating crop at heading.
  \item \emph{Crop3} (E half, $x \in [0,500]\,\text{m}$): E--W rows;
        intermediate height and attenuation.
\end{itemize}
Four metal farm buildings and one diagonal irrigation canal ($45^\circ$, $900 \times 10\,\text{m}$) are included to introduce a realistic radio environment.
The material parameters for all structures follow ITU-R P.527-4~\cite{itu_p527}
(dry soil: $\epsilon_r\!=\!3.0$, $\sigma\!=\!0.001\,\text{S/m}$;
canal water: $\epsilon_r\!=\!80$, $\sigma\!=\!0.01\,\text{S/m}$). 
We assume a private 5G NR deployment where IoT sensor nodes operate as NR RedCap (Reduced Capability) UEs as defined in 3GPP Release 17 \cite{3gpp_tr38875}. The RSRP threshold  $\gamma$= -85 dBm is consistent with minimum NR coverage requirements for RedCap devices in outdoor deployments.

\subsection{Seasonal Channel Model}

We define four discrete growth stages $\stageset =
\{\textit{sowing},\, \textit{vegetative},\, \textit{heading},\,
\textit{harvest}\}$, each with stage-specific vegetation permittivity
$\epsilon_r^{(s)}$, conductivity $\sigma_\text{veg}^{(s)}$, and canopy
height $h^{(s)}$ per crop zone, derived from ITU-R P.833-9~\cite{itu_p833}. \emph{Crop1} and \emph{Crop3} follow similar seasonal trajectories, while \emph{Crop2} represents the worst-case attenuation during the mature stages due to its taller canopy. The parameters for the three crops are listed in Table~\ref{tab:crop_params}.

\begin{table}[t]
\centering
\caption{Multi-crop seasonal propagation parameters (ITU-R P.833).
$\varepsilon_r$ (relative permittivity), $\sigma$ (S/m), $h$ (m)}
\label{tab:crop_params}
\setlength{\tabcolsep}{4pt}
\begin{tabular}{llccc}
\toprule
\textbf{Crop} & \textbf{Stage} & $\varepsilon_r$ & $\sigma$ (S/m) & $h$ (m) \\
\midrule
\multirow{4}{*}{Crop1 (NW)}
  & Sowing     & 3.0 & 0.001 & 0.0 \\
  & Vegetative & 4.0 & 0.005 & 0.4 \\
  & Heading    & 4.6 & 0.010 & 0.8 \\
  & Harvest    & 3.0 & 0.001 & 0.1 \\
\midrule
\multirow{4}{*}{Crop2 (SW)}
  & Sowing     & 3.0 & 0.001 & 0.0 \\
  & Vegetative & 4.5 & 0.008 & 1.0 \\
  & Heading    & 5.0 & 0.020 & 2.5 \\
  & Harvest    & 3.2 & 0.002 & 0.2 \\
\midrule
\multirow{4}{*}{Crop3 (E)}
  & Sowing     & 3.0 & 0.001 & 0.0 \\
  & Vegetative & 3.8 & 0.005 & 0.3 \\
  & Heading    & 4.0 & 0.015 & 1.2 \\
  & Harvest    & 2.5 & 0.001 & 0.1 \\
\bottomrule
\end{tabular}
\end{table}
\subsection{Coverage Metric}
Let $K$ base stations (BSs) be deployed in the farm, with locations  $ \mathcal{P} = \{\mathbf{p}_1,\mathbf{p}_2,\ldots,\mathbf{p}_K\}$, where $\mathbf{p}_j = (x_j,y_j)$ denotes the position of BS $j$ in the farm area. The height (z-axis) of all BSs is set to a fixed value $h_{BS}$ and UEs are assumed to be located within the crop canopy at height $h_{\text{UE}}$. For a UE location $\mathbf{q}=(x,y)$ and growth stage $s \in \stageset$, let $P_r^{(j)}(\mathbf{q};s)$ denote the received power from BS $j$. Assuming strongest-BS association, the best-server received power at location $\mathbf{q}$ is
\begin{equation}
P_r^\star(\mathbf{q};s)
= \max_{j=1,\ldots,K} P_r^{(j)}(\mathbf{q};s).
\end{equation}

A location $\mathbf{q}$ is considered \emph{covered} if the received power exceeds the minimum Reference Signal Received Power (RSRP) threshold $\gamma$ required 
for reliable NR IoT connectivity, i.e., $ P_r^\star(\mathbf{q};s) \ge \gamma $. 
Let $\eta_s(\mathcal{P})$ denote the coverage fraction at stage $s$
for placement $\mathcal{P}$ as defined in \eqref{eq:cov}. The stage-$s$ coverage fraction for a BS placement $\mathcal{P}$ is defined as
\begin{equation}
\eta_s(\mathcal{P})
=
\frac{1}{|\mathcal{Q}|}
\sum_{\mathbf{q}\in\mathcal{Q}}
\mathbf{1}\!\left[P_r^\star(\mathbf{q};s)\ge\gamma\right],
\label{eq:cov}
\end{equation}
where $\mathbf{1}[\cdot]$ is the indicator function.

\subsection{Problem Formulation}

Due to seasonal vegetation growth, a placement optimal for one 
stage may perform poorly in another. We therefore adopt a robust 
criterion that maximizes the worst-case coverage fraction across 
all stages:
\begin{equation}
    \max_{\mathcal{P}} \min_{s \in \mathcal{S}} \eta_s(\mathcal{P}),
    \label{eq:maximin}
\end{equation}
where the objective ensures reliable connectivity even under the 
most adverse propagation conditions.



In practice, estimating $\eta_s(\mathcal{P})$ through field
measurements would require extensive drive tests across the farm. In this work, we instead compute coverage maps using
ray tracing of the farm scene under the stage-$s$ channel parameters. As mentioned, each evaluation of the objective in \eqref{eq:maximin} requires multiple expensive ray-tracing simulations. Consequently, a single objective evaluation requires a full ray-tracing simulation, making exhaustive search over the continuous BS location space computationally infeasible. In the next section, we therefore propose a Bayesian optimization framework to efficiently solve \eqref{eq:maximin} using a small number of ray-tracing evaluations.
\section{Bayesian Optimisation Framework}
\label{sec:bo}
The seasonal-robust placement problem in \eqref{eq:maximin} requires maximizing the worst-case coverage fraction across all
growth stages. For convenience, we define the scalar objective \begin{equation}
f(\mathcal{P}) = \min_{s \in \stageset} \eta_s(\mathcal{P}),
\end{equation} where $\eta_s(\mathcal{P})$ denotes the stage-$s$ coverage
fraction defined in \eqref{eq:cov}. Evaluating $f(\mathcal{P})$ requires computing coverage maps for all seasonal stages via ray tracing of the farm scene, making each objective evaluation computationally expensive. To efficiently search the continuous BS placement space, we adopt BO, which is well suited for optimizing expensive black-box functions \cite{greenhill2020bayesian}. BO iteratively
constructs a probabilistic surrogate model of the objective function using previously evaluated placements and selects new evaluation points by maximizing an acquisition function that balances exploration and exploitation.

For optimization, the BS placement $\mathcal{P}$ is parameterized by a vector $\mathbf{x} \in [0,1]^{2K}$ containing the normalized $(x,y)$ coordinates of the $K$ base stations. 
The BO procedure alternates between fitting a Gaussian-process surrogate to the observed objective values and selecting new placements via an acquisition function.

\subsection{Gaussian-Process Surrogate}

As evaluating $f(\mathbf{x})$ requires multiple ray-tracing simulations, we approximate the objective using a GP surrogate. Let $\mathcal{D}_n =
\{(\mathbf{x}_i, f_i)\}_{i=1}^n$ denote the set of evaluated placements and their corresponding objective values $f_i = f(\mathbf{x}_i)$. We model the objective as

\begin{equation}
f(\mathbf{x}) \mid \mathcal{D}_n
\;\sim\;
\GP\!\bigl(\mu_n(\mathbf{x}),\, \sigma_n^2(\mathbf{x})\bigr).
\end{equation}

The GP is fitted using exact inference in GPyTorch~\cite{gardner2018gpytorch} with a Mat\'{e}rn-$5/2$ kernel with automatic relevance determination (ARD). The used kernel provides a flexible model for continuous objectives that may exhibit moderate spatial irregularities due to multipath propagation \cite{williams2006gaussian}. ARD length-scales allow the surrogate to automatically identify which BS coordinates most strongly influence the worst-case coverage. For example, BSs located near buildings or dense crop
zones may exhibit shorter learned length-scales, indicating higher sensitivity of the maximin objective to their position. Since the ray-tracing simulator is deterministic, we employ a near-zero observation noise model to stabilize GP training.

\subsection{Acquisition and Optimization Loop}

To select the next placement to evaluate, we adopt the Upper Confidence Bound (UCB) acquisition  function~\cite{srinivas2010gaussian} $\alpha_{\text{UCB}}(\mathbf{x}) = \mu(\mathbf{x}) + \sqrt{\beta} \sigma(\mathbf{x}),$ where $\mu(\mathbf{x})$ and $\sigma(\mathbf{x})$ are the posterior mean and standard deviation of GP, and $\beta > 0$ is a trade-off parameter controlling exploration versus exploitation. We use a fixed $\beta = 2.0$ in all iterations, which is a standard practical choice \cite{srinivas2010gaussian}. UCB is well suited for expensive black-box optimization because it explicitly balances sampling in high-mean regions (exploitation) and high-uncertainty regions (exploration), and its linear form in the GP posterior makes it efficient to maximize. The acquisition function is maximized using L-BFGS-B with 20 random restarts within the bounded domain $[0,1]^{2K}$ using the BoTorch framework~\cite{balandat2020botorch}.
The full BO procedure is summarized in Algorithm~\ref{alg:bo}. The optimization begins with $n_{\text{seed}}$ Latin-hypercube (LHS) seed evaluations to initialize the surrogate model, followed by $n_{\text{BO}}$ BO iterations guided by UCB, for a total budget of $n_{\text{seed}}+n_{\text{BO}}$ RT evaluations. Note that the returned solution $\mathbf{x}^\star$ is the best directly observed placement in $\mathcal{D}$ rather than the GP posterior mean maximizer.

\begin{algorithm}[t]
  \caption{GPBO for maximin seasonal BS placement}
  \label{alg:bo}
  \begin{algorithmic}[1]
    \STATE Sample $\mathbf{X}_0 \sim \text{LHS}([0,1]^{2K},\; n_{\text{seed}})$
    \STATE Evaluate $f_i \leftarrow \min_{s} \eta_s(\mathbf{X}_0[i])$ via RT
    \STATE $\mathcal{D} \leftarrow \{(\mathbf{X}_0, \mathbf{f}_0)\}$
    \FOR{$t = 1, \ldots, n_{\text{BO}}$}
      \STATE Fit GP on $\mathcal{D}$ to obtain $(\mu_t, \sigma_t)$
      \IF{$\max_{\mathbf{x}} \alpha_{\text{UCB}}(\mathbf{x}) < 0.005$}
        \STATE \textbf{break} \quad (convergence)
      \ENDIF
      \STATE $\mathbf{x}_t \leftarrow \arg\max_{\mathbf{x}} \alpha_{\text{UCB}}(\mathbf{x})$
      \STATE Evaluate $f_t \leftarrow \min_{s \in \stageset} \eta_s(\mathbf{x}_t)$ via RT
      \STATE $\mathcal{D} \leftarrow \mathcal{D} \cup \{(\mathbf{x}_t, f_t)\}$
    \ENDFOR
    \STATE $\mathbf{x}^\star = \arg\max_{\mathbf{x}_i \in \mathcal{D}} f_i$
    \RETURN $\mathbf{x}^\star$ 
  \end{algorithmic}
\end{algorithm}
\section{Numerical Results}
\label{sec:results}
We evaluate the proposed BO framework for robust BS placement in the agricultural IoT scenario  (Section~\ref{sec:sysmod}). Results focus on: (i) optimization efficiency of BO compared with baseline methods and  comparison of different acquisition strategies, and (ii) seasonal robustness of the resulting deployment.

\subsection{Simulation Setup}
All experiments use the farm scene described in Section~\ref{sec:sysmod}. We optimize $K=3$ base stations, resulting in a $6$-dimensional search space corresponding to $(x,y)$ coordinates of three BSs. The optimization domain is normalized to $[0,1]^6$ and mapped to the physical farm region $[-500,500]^2$ before ray-tracing evaluation. Coverage maps are generated using the Sionna RT ray-tracing engine at carrier frequency $f_c = 3.5$~GHz. The proposed GPBO framework is frequency-agnostic and can be directly applied to sub-GHz bands for LPWAN-based deployments. Each BS transmits with power $P_{\mathrm{tx}} = 33$~dBm using a 3GPP TR~38.901 sector antenna pattern with $5^\circ$ electrical downtilt. The BS height is fixed to $20$~m, while IoT sensor nodes are modelled as UEs located within the crop canopy at height $0.3$~m. The ray tracer launches $10^7$ rays per BS with a maximum interaction depth of $15$, generating an RSRP map over a $1000\times1000$ grid covering the entire farm. The primary performance metric is the coverage fraction defined in~\eqref{eq:cov}, with an RSRP threshold $\gamma=-85\dBm$, $n_{\text{seed}}=10$  and $n_{\text{BO}}=40$. Each evaluation of the objective $f(\mathbf{x})$ requires computing coverage maps for all four seasonal stages defined in  Section~\ref{sec:sysmod}, resulting in four simulations of ray-tracing per candidate placement. We compare four deployment strategies: (1)~\textbf{Centre placement}, a heuristic placing BSs near the geometric centre of each half-field; (2)~\textbf{Budget-matched random search}, sampling candidate placements uniformly at random using the same RT evaluation budget as BO; (3)~\textbf{Particle swarm optimization (PSO)}, a population-based metaheuristic with $N_p = 10$ particles initialized from the top-$N_p$ LHS seeds and run for $\lfloor n_\text{BO}/N_p \rfloor$ iterations, matching BO's RT budget exactly; and (4)~\textbf{Robust BO (proposed)}, the GPBO algorithm described in Section~\ref{sec:bo} with UCB ($\beta = 2.0$) as the default acquisition function. To justify this choice, we additionally compare UCB against logEI~\cite{ament2023unexpected}, Thompson sampling~\cite{daniel2018tutorial}, and MES~\cite{wang2017max}, all initialized from the same LHS seed with identical RT budgets.





\begin{figure}[t]
  \centering  \includegraphics[width=8.5cm, height=4cm]{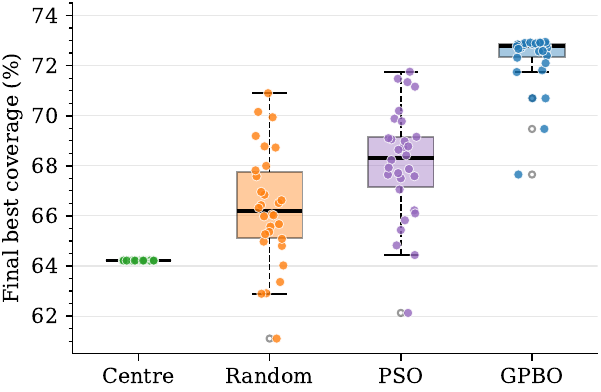}
  \caption{Distribution of final best coverage across 30 runs. Each method is initialized with the same LHS seeds per run. The Centre placement is deterministic as the BSs are always placed at the same fixed geometric locations regardless of the run.}
  \label{fig:boxplot}
\end{figure}

\subsection{Optimization Convergence}

Fig.~\ref{fig:boxplot} shows the distribution of final best worst-case coverage across 30 independent runs. The proposed GPBO achieves the highest median coverage of 72.8\% with a low inter-quartile range (IQR) of 0.1~pp, indicating stable and consistent optimization behavior. PSO (68.1\%) and random search (66.0\%) improve over the deterministic centre placement (63.7\%), but remain substantially inferior to GPBO. The zero spread of the centre placement boxplot is expected, as the BS locations are fixed geometric positions independent of the run.

\begin{figure}[t]
  \centering  \includegraphics[width=8.5cm, height=5cm]{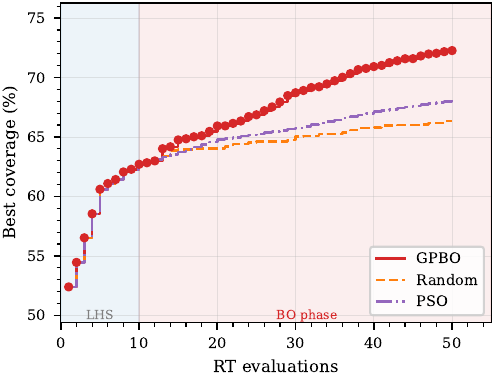}
  \caption{Best-so-far maximin coverage vs.\ RT evaluations for GPBO, Random and PSO. }
  \label{fig:convergence}
\end{figure}

Fig.~\ref{fig:convergence} shows the median best-so-far worst-case coverage as a function of RT evaluations. After the $n_\text{seed} = 10$ LHS seed evaluations, the median coverage across all methods is approximately 62.3\%. GPBO then rapidly improves, reaching high-quality worst-case coverage of 72.8\% within 50 RT evaluations, demonstrating superior sample efficiency over the baselines. PSO improves more slowly, reaching 68.1\%, as the limited number of swarm iterations constrains effective exploration of the 6-dimensional placement space. Random search, lacking any surrogate model, shows the slowest and most irregular convergence.

Among the four acquisition functions evaluated, i.e., UCB, logEI, MES, and Thompson Sampling, UCB and logEI perform comparably, both reaching a median worst-case coverage of approximately 72.8\% after 50 RT evaluations. MES remains competitive but converges to a slightly lower median of approximately 71.0\%. Thompson Sampling underperforms at approximately 67.0\%, likely due to its stochastic sampling nature combined with the limited RT evaluation budget. UCB is adopted as the default acquisition function for its consistently strong performance and computational simplicity.

\begin{figure*}[t]
  \centering
\includegraphics[width=\textwidth,height=4.0cm]{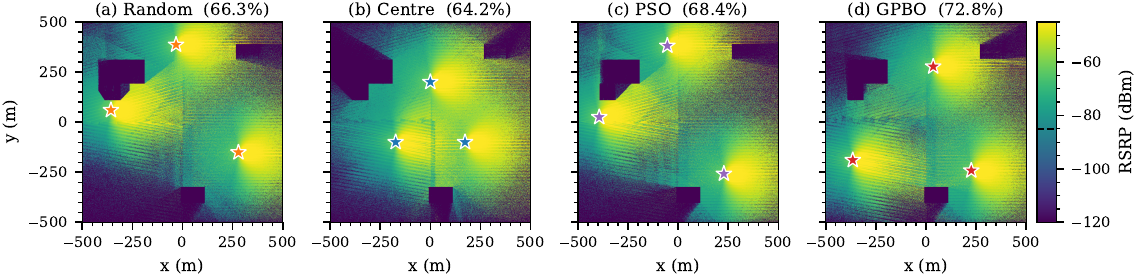}
  \caption{Per-cell RSRP maps (dBm) computed via Sionna ray tracing at heading stage for four placements: (a) Budget-matched random search, (b) geometric centres, (c) Particle swarm optimization and (d) GPBO. Stars denote BS locations; dashed colorbar line marks the -85 dBm threshold.}
  \label{fig:scene_rt}
\end{figure*}

\begin{table}[t]
  \centering
  \caption{Coverage fraction (\%) per growth stage (median [IQR]) across 30 runs.}
  \label{tab:stage_results}
  \renewcommand{\arraystretch}{1.15}
  \begin{tabular}{lcccc}
    \toprule
    \textbf{Method} & \textbf{Sowing} & \textbf{Vegetative} & \textbf{Heading} & \textbf{Harvest} \\
    \midrule
    Centre          & 77.7 [0.0] & 65.5 [0.0] & 63.7 [0.0] & 77.2 [0.0] \\
    Random          & 79.6 [3.9] & 66.5 [4.1] & 66.0 [3.9] & 79.0 [4.2] \\
    PSO             & 80.6 [3.4] & 68.1 [3.6] & 68.1 [3.3] & 80.1 [3.6] \\
    GPBO       & 84.7 [0.2] & 72.8 [0.1] & 72.7 [0.1] & 84.2 [0.2] \\
    \bottomrule
  \end{tabular}
\end{table}

\subsection{Seasonal Robustness}

Fig.~\ref{fig:scene_rt} shows the RT-computed RSRP maps at the heading stage, which is the most challenging growth stage due to maximum canopy height and vegetation attenuation for all four placement methods. GPBO places the three BSs to achieve notably more uniform coverage across all three crop zones compared to the baselines, which leave significant portions of the farm below the $-85$~dBm threshold. Table~\ref{tab:stage_results} quantifies this improvement across all growth stages as median [IQR] over 30 runs. The heading stage is the binding constraint for all methods, confirming it as the worst-case season. GPBO achieves 72.7\%  median heading-stage coverage, outperforming PSO by $+4.6$~pp, random search by 
$+6.6$~pp, and centre placement by $+8.9$~pp. Importantly, the gains of GPBO extend across all growth stages, confirming that the proposed formulation does not sacrifice non-worst-case performance to protect against the hardest stage.

\section{Conclusion}
\label{sec:conclusion}

We presented a Bayesian optimization framework for seasonal-robust BS placement in multi-crop agricultural IoT deployments. By formulating the placement objective as a maximin worst-case coverage over four crop-growth stages, we reduce an otherwise intractable seasonal robustness problem to a standard single-output GPBO task.
In a \SI{1}{\km\squared} farm scene at \SI{3.5}{\GHz}, with three different crop zones, the proposed scheme achieves high-quality worst-case coverage in fewer than $50$ RT evaluations which outperforms budget-matched PSO, random search and centre placement across all crop growth stages. Future work will validate the RT predictions against field RSRP measurements across crop stages to quantify the accuracy of the model and extend the presented discrete stage problem to a continuous-time crop seasonal model.

\bibliographystyle{IEEEtran}
\bibliography{spcom2026_seasonal}

@article{elijah2018overview,
  title={An overview of {Internet of Things (IoT)} and data analytics in agriculture: Benefits and challenges},
  author={Elijah, Olakunle and Rahman, Tharek Abdul and Orikumhi, Igbafe and Leow, Chee Yen and Hindia, MHD Nour},
  journal={{IEEE Internet of things Journal}},
  volume={5},
  number={5},
  pages={3758--3773},
  year={2018},
  publisher={IEEE}
}

@article{majone2013wireless,
  title={Wireless sensor network deployment for monitoring soil moisture dynamics at the field scale},
  author={Majone, Bruno and Viani, Federico and Filippi, E and Bellin, Alberto and Massa, Andrea and Toller, G and Robol, Fabrizio and Salucci, Marco},
  journal={Procedia environmental sciences},
  volume={19},
  pages={426--435},
  year={2013},
  publisher={Elsevier}
}

@article{garcia2021deployment,
  title={Deployment strategies of soil monitoring WSN for precision agriculture irrigation scheduling in rural areas},
  author={Garc{\'\i}a, Laura and Parra, Lorena and Jimenez, Jose M and Parra, Mar and Lloret, Jaime and Mauri, Pedro V and Lorenz, Pascal},
  journal={Sensors},
  volume={21},
  number={5},
  pages={1693},
  year={2021},
  publisher={Multidisciplinary Digital Publishing Institute}
}

@inproceedings{yangyang2004particle,
  title={Particle swarm optimization for base station placement in mobile communication},
  author={Yangyang, Zhang and Chunlin, JI and Ping, Yuan and Manlin, LI and Chaojin, Wang and Guangxing, Wang},
  booktitle={IEEE International Conference on Networking, Sensing and Control, 2004},
  volume={1},
  pages={428--432},
  year={2004},
  organization={IEEE}
}

@inproceedings{rahim2017foliage,
  title={Foliage attenuation measurement at millimeter wave frequencies in tropical vegetation},
  author={Rahim, Hairani Maisarah and Leow, Chee Yen and Abd Rahman, Tharek and Arsad, Arsany and Malek, Muhammad Arif},
  booktitle={2017 IEEE 13th Malaysia International Conference on Communications (MICC)},
  pages={241--246},
  year={2017},
  organization={IEEE}
}

@article{majumdar2025enhancing,
  title={Enhancing sustainable {5G} powered agriculture 4.0: Summary of low power connectivity, internet of {UAV} things, {AI} solutions and research trends},
  author={Majumdar, Parijata and Mitra, Sanjoy and Bhattacharya, Diptendu and Bhushan, Bharat},
  journal={Multimedia tools and applications},
  volume={84},
  number={17},
  pages={17389--17433},
  year={2025},
  publisher={Springer}
}

@techreport{3gpp_tr38875,
  author       = "{3rd Generation Partnership Project ({3GPP})}",
  title        = "Study on Support of Reduced Capability {NR} Devices",
  institution  = "{3GPP}",
  type         = "Technical Report",
  number       = "TR 38.875",
  note         = "Release 17",
  year         = "2021",
}

@article{hurley2002planning,
  title={Planning effective cellular mobile radio networks},
  author={Hurley, Stephen},
  journal={IEEE Transactions on Vehicular Technology},
  volume={51},
  number={2},
  pages={243--253},
  year={2002},
  publisher={IEEE}
}

@inproceedings{valavanis2014base,
  title={Base-station location optimization for LTE systems with genetic algorithms},
  author={Valavanis, Ioannis K and Athanasiadou, Georgia and Zarbouti, Dimitra and Tsoulos, George V},
  booktitle={European Wireless 2014; 20th European Wireless Conference},
  pages={1--6},
  year={2014},
  organization={VDE}
}

@inproceedings{wu2024multiobjective,
  title={Multiobjective {Bayesian} optimization for antenna placement in in-building distributed antenna system},
  author={Wu, Xilei and Huang, Pei-Qiu and Song, Linqi and Liu, Hai-Lin and Zhang, Qingfu},
  booktitle={2024 IEEE Congress on Evolutionary Computation (CEC)},
  pages={1--8},
  year={2024},
  organization={IEEE}
}

@article{sato2024bayesian,
  title={Bayesian optimization framework for channel simulation-based base station placement and transmission power design},
  author={Sato, Koya and Suto, Katsuya},
  journal={IEEE Networking Letters},
  volume={6},
  number={4},
  pages={217--221},
  year={2024},
  publisher={IEEE}
}

@misc{hoydis2023sionna,
  author        = {J. Hoydis and S. Cammerer and F. A. Aoudia and A. Vem
                   and N. Binder and G. Marcus and A. Keller},
  title         = {{Sionna RT}: Differentiable Ray Tracing for Radio
                   Propagation Modeling},
  howpublished  = {arXiv:2303.11103},
  year          = {2023}
}

@techreport{itu_p833,
  institution  = {ITU-R},
  title        = {Attenuation in Vegetation},
  number       = {Recommendation P.833-9},
  year         = {2016}
}

@techreport{itu_p527,
  institution  = {ITU-R},
  title        = {Electrical Characteristics of the Surface of the {Earth}},
  number       = {Recommendation P.527-5},
  year         = {2019}
}

@article{greenhill2020bayesian,
  title={Bayesian optimization for adaptive experimental design: A review},
  author={Greenhill, Stewart and Rana, Santu and Gupta, Sunil and Vellanki, Pratibha and Venkatesh, Svetha},
  journal={IEEE access},
  volume={8},
  pages={13937--13948},
  year={2020},
  publisher={IEEE}
}

@article{daniel2018tutorial,
  title={A tutorial on thompson sampling},
  author={Daniel, J Russo and Benjamin, Van Roy and Abbas, Kazerouni and Ian, Osband and Zheng, Wen},
  journal={Foundations and Trends{\textregistered} in Machine Learning},
  volume={11},
  number={1},
  pages={1--99},
  year={2018},
  publisher={Emerald Publishing Limited}
}

@inproceedings{wang2017max,
  title={Max-value entropy search for efficient {Bayesian} optimization},
  author={Wang, Zi and Jegelka, Stefanie},
  booktitle={{International Conference on Machine Learning}},
  pages={3627--3635},
  year={2017},
  organization={PMLR}
}

@inproceedings{gardner2018gpytorch,
  author    = {J. Gardner and G. Pleiss and K. Q. Weinberger
               and D. Bindel and A. G. Wilson},
  title     = {{GPyTorch}: Blackbox Matrix-Matrix {Gaussian} Process
               Inference with {GPU} Acceleration},
  booktitle = {Proc.\ NeurIPS},
  year      = {2018}
}

@inproceedings{balandat2020botorch,
  author    = {M. Balandat and B. Karrer and D. R. Jiang and S. Daulton
               and B. Letham and A. G. Wilson and E. Bakshy},
  title     = {{BoTorch}: A Framework for Efficient {Monte Carlo}
               {Bayesian} Optimization},
  booktitle = {Proc.\ NeurIPS},
  year      = {2020}
}

@inproceedings{ament2023unexpected,
  author    = {S. Ament and M. Daulton and D. Eriksson
               and M. Balandat and E. Bakshy},
  title     = {Unexpected Improvements to Expected Improvement for
               {Bayesian} Optimization},
  booktitle = {Proc.\ NeurIPS},
  year      = {2023}
}

@inproceedings{srinivas2010gaussian,
  author    = {N. Srinivas and A. Krause and S. M. Kakade and M. Seeger},
  title     = {{Gaussian} Process Optimization in the Bandit Setting:
               No Regret and Experimental Design},
  booktitle = {Proc.\ ICML},
  year      = {2010}
}

@book{williams2006gaussian,
  title={Gaussian processes for machine learning},
  author={Williams, Christopher KI and Rasmussen, Carl Edward},
  volume={2},
  number={3},
  year={2006},
  publisher={MIT press Cambridge, MA}
}

\end{document}